\documentclass[12pt]{iopart}
\usepackage{graphicx}
\usepackage{bm}
\usepackage[mathscr]{eucal}

\newcommand{\LB}{\left[}
\newcommand{\RB}{\right]}
\newcommand{\aee}{a_{\mbox{\scriptsize e}}}      
\newcommand{\aen}{a_{\mbox{\scriptsize n}}}      
\newcommand{\anot}{\alpha_0}                     
\newcommand{\Enot}[1]{E_0^{(#1)}}                
\newcommand{\Eret}{E_{\mbox{\scriptsize r}}}     
\newcommand{\sE}{\mathscr{E}}                    
\newcommand{\Ip}{I_{\mbox{\scriptsize p}}}       
\newcommand{\lnt}{\lambda_0}                     
\newcommand{\onot}{\omega_0}                     
\newcommand{\op}{\omega_{\mbox{\scriptsize p}}}  
\newcommand{\sw}{\sigma_{\mbox{\scriptsize w}}}  
\newcommand{\ten}[1]{\times 10^{#1}}             
\newcommand{\tret}{t_{\mbox{\scriptsize r}}}     
\newcommand{\Tw}{T_{\mbox{\scriptsize w}}}       
\newcommand{\Up}{U_{\mbox{\scriptsize p}}}       
\newcommand{\Ve}{V_{\mbox{\scriptsize e}}}       
\newcommand{\Vn}{V_{\mbox{\scriptsize n}}}       
\newcommand{\xcap}{x_{\mbox{\scriptsize cap}}}  

\newcommand{\dg}{$^{\circ}$ }

\begin{document}

\title{Plasmon signatures in high harmonic generation}
\author{J Zanghellini\dag, Ch Jungreuthmayer\ddag\ and
  T Brabec\ddag}

\address{\dag\ Institute of Chemistry, Karl-Franzens-University Graz,\\
  Heinrichstra{\ss}e 28, A-8010 Graz, Austria, EU}
\address{\ddag\ Center for Photonics Research, University of Ottawa,\\
  150 Louis Pasteur, Ottawa, ON, K1N 6N5 Canada}
\eads{\mailto{juergen.zanghellini@uni-graz.at}, \mailto{brabec@uottawa.ca}}

\date{\today}
\submitto{J. Phys. B (2005), accepted.}

\begin{abstract}
High harmonic generation in polarizable multi-electron systems is investigated in the framework of
multi-configuration time-dependent Hartree-Fock. The harmonic spectra exhibit two cut offs. The first
cut off is in agreement with the well established, single active electron cut off law. The second cut
off presents a signature of multi-electron dynamics. The strong laser field excites non-linear plasmon
oscillations. Electrons that are ionized from one of the multi-plasmon states and recombine to the
ground state gain additional energy, thereby creating the second plateau.
\end{abstract}

\maketitle

When an intense laser pulse is focused onto a noble gas jet, high harmonic generation (HHG) takes place.
High harmonic radiation is created in a three step process \cite{corkum93}. The valence electron is set
free by tunnel ionization. In the continuum, the electron
is accelerated and follows the quiver motion of
the laser field. When the laser field changes sign, the electron is driven back towards the parent ion.
Finally, the electron recombines to the ground state upon recollision, and an xuv photon is emitted.

The theory of HHG is based on the single-active-electron (SAE) approximation \cite{krause92}, assuming that
only the valence electron interacts with the strong laser field while the residual electron core remains
frozen. The valence electron and the core electrons are regarded as uncorrelated. HHG has been performed
with noble gas atoms and clusters \cite{tisch97}, and with small molecules \cite{liang94,nalda04}. All
experiments were found to be in agreement with SAE theory.

Experimental \cite{lezius02,bhardwaj03,misha03} and theoretical \cite{kitzler04} evidence was found that SAE
theories cannot describe optical field ionization of highly polarizable systems, such as large molecules and
metallic clusters. Due to the high electron mobility and polarizability, a factorization into valence and
core electrons is no longer valid and the complete, correlated multi-electron (ME) dynamics has to be taken
into account. This raises the question as to which extent the SAE approximation is applicable to non-perturbative
phenomena in complex materials \cite{veniard01,bandrauk04}.

In this article we investigate HHG in highly polarizable molecules by an one-dimensional (1D) multi-configuration
time-dependent Hartree-Fock (MCTDHF) analysis. MCTDHF is a recently developed method allowing to account for the
electron correlation in a numerically converged manner \cite{zanghellini03,zanghellini04,kato04,caillat05,nest05}.
Our analysis reveals that in contrast to HHG in noble gases, where the harmonic spectrum exhibits one plateau and
cut off, a second cut off is identified, extending far beyond the standard cut off. This second cut off originates
from the ME nature of the bound electrons. The strong laser field excites non-linear, collective electron
oscillations and populates multi-plasmon states that oscillate at a multiple of the plasmon frequency. The second
plateau is generated by electrons that ionize from the multi-plasmon states and recombine to the ground state.
The energy difference between excited and ground state determines the difference between first and second cut off.

\section{The MCTDHF method}
Here, we demonstrate the general idea of the MCTDHF-ansatz by means of an example, containing two 1D particles.
For simplicity we will not take spin into consideration, although it is included in the calculation presented
below. For an extensive review of the MCTDHF theory and formalism we refer to \cite{caillat05} and references
therein.

MCTDHF makes the ansatz
\begin{eqnarray}
  \label{eq:ansatz}
  \Psi (x,y;t)= \frac{1}{\sqrt{2}}\sum_{j_1<j_2}^m A_{j_1j_2}(t)
                \LB \varphi_{j_1}(x;t)\varphi_{j_2}(y;t)-\varphi_{j_2}(x;t)\varphi_{j_1}(y;t)\RB
\end{eqnarray}
Thus MCTDHF consists in approximating the exact wave function as linear combination of ${m \choose 2}$ different
Slater determinants. Since the coefficients and the $m$ linearly independent expansion functions are time-dependent,
an additional constraint is needed. Without loss of generality, it is most convenient to impose ortho-normality on
the expansion functions, i.e. $\left\langle \varphi_i(t) | \varphi_j(t)\right\rangle=\delta_{i,j}$. Additionally,
we impose $\left\langle \frac{\rmd}{\rmd t}\varphi_i(t) | \varphi_j(t)\right\rangle= 0$ to uniquely define the
expansion in \eref{eq:ansatz} \cite{caillat05}.

To derive equations of motion the Dirac-Frenkel variational principle \cite{dirac30,frenkel34},
$\left\langle \delta\Psi \left| \rmi \frac{\rmd}{\rmd t} -H(t) \right| \Psi\right\rangle= 0$, is imposed. Here,
$H$ denotes the time-dependent Hamiltonian describing the electronic system. The such derived equations are of
the form
\begin{eqnarray}
  \rmi \frac{\rmd}{\rmd t} \bi{A}= F[\boldsymbol{\varphi}] \bi{A}\\
  \rmi \frac{\rmd}{\rmd t} \boldsymbol{\varphi}=
       G[\bi{A},\boldsymbol{\varphi}] \boldsymbol{\varphi}.
\end{eqnarray}
The time evolution is obtained by applying non-linear operators $F$ and $G$ on vectors
$\bi{A}= (A_{1,2},...,A_{n-1,n})^T$ and $\boldsymbol{\varphi}=(\varphi_1,...,\varphi_m)^T$, respectively. For
details on the equations see \cite{caillat05}.

\begin{figure}[t]
\centerline{\includegraphics[width=0.5\columnwidth]{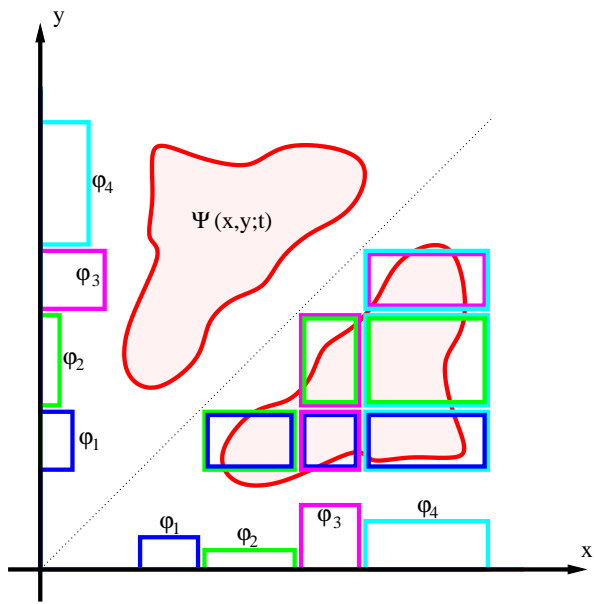}}
\caption{Approximation of a 2-dimensional, correlated wave function $\Psi (x,y;t)$ as a sum of 6 determinants,
indicated as rectangular patches. The expansion functions $\varphi_i$ belonging to their corresponding determinant
are drawn along the axes in their respective color.
\label{fig:mctdhf}}
\end{figure}
The principle of MCTDHF can be seen in figure~\ref{fig:mctdhf}. The fully correlated wave function $\Psi (x,y;t)$
is approximated by 6 Slater determinants, implying 4 expansion functions per particle. Thus the wave packet is
reassembled in a kind of "patch work". Each patch represents one single Slater determinant. In
figure~\ref{fig:mctdhf} we have, for simplicity, assumed the single electron orbitals to be of rectangular shape.
However, we emphasize that generally, the expansion orbitals will be a priori unknown. It is important to note
that not only the expansion coefficients evolve with time, but the single particle orbital, too. Hence MCTDHF may
be interpreted as a truncated configuration-interaction (CI) expansion, in which both, coefficients, $A_{j_1j_2}(t)$,
and orbitals, $\varphi_j(x;t)$, are optimized. For every time step and fixed number of configuration an optimal
expansion is warranted by the Dirac-Frenkel variational principle. Thus resulting in a compact representation of
the wave function and hence compressing the necessary storage amount.

MCTDHF is applicable to more complex systems because already with "small" configuration numbers the "essential
physics" is covered, i.e. the number of physically important expansion orbitals, is always much smaller than the
number of time-independent basis functions in conventional approaches. Therefore MCTDHF scales more slowly,
allowing the treatment of small molecules beyond state of the art 2-electrons-calculation.

By increasing the number of configurations MCTDHF allows a systematic inclusion of correlation, converging to the
exact solution for $m\rightarrow\infty$.

These advantages are achieved at the expense of linearity and locality, since the evolution equations in MCTDHF are
both, non-linear and non-local. Just as time-dependent Hartree-Fock (TDHF), also MCTDHF with a finite number of
configurations suffers, in principle, from the problem of non-linearity of the evolution equations which may lead
to a violation of the superposition principle. However, in the case of MCTDHF this problem is greatly reduced by
adding additional configurations.

The ability of MCTDHF to correctly describe the correlated dynamics of electrons has been assessed in recent
studies \cite{zanghellini04,kato04} using two-electron systems.

Here, we report calculations for $n=4$ electrons using $m=8$ expansion functions. The Schr\"odinger equation is
solved in a simulation box with size $l=\pm360$\,at.u. and on a uniform 1D-grid with 2400 grid points, using a
second order finite difference representation. To avoid reflection at the boundaries, complex absorption potentials
(CAP) are used. That is, the total Hamiltonian is modified by adding $\rmi\sum_i^n\{ 1-\cos\LB\pi (x_i-\xcap )/
(|l|-\xcap )\RB\}$ for $|x_i|>\xcap=270$\,at.u. and $0$ otherwise. The time-integration is performed by a
self-adaptive, high-order Runge-Kutta integrator with a relative numerical accuracy of $10^{-8}$. Convergency was
checked with respect to all of these parameters. In particular, increasing the number of expansion functions to
$m=12$ does not change our finding and changes for instance the ionization yield by less than 4\%, indicating that
our calculations are essentially converged.

\section{High harmonic generation in polarizable molecules}

The 1D MCTDHF analysis is based on the solution of the 1D Schr\"odinger equation for the $n=4$-electron potential
$V = \sum_{i=1}^n \LB-\Vn(x_i)+ \sum_{j>i}^n \Ve(x_i-x_j)\RB$. Here, $\Vn = Z / \sqrt{x_i^2+\aen^2}$ refers to
the nuclear binding potential, $Z$ is the charge state, and $\aen$ is the shielding parameter of the
electron-nucleus interaction. The shielded model potential represents an atom or a small cluster. We believe
that it is closer to a small cluster, with a harmonic oscillator potential part close to the center, and a
Coulomb far-range potential far away from the center. Further, $\Ve = 1 / \sqrt{(x_i-x_j)^2 + \aee^2}$ represents
the electron-electron interaction potential with shielding parameter $\aee$. The laser is coupled in velocity gauge
and in dipole approximation. Atomic units are used throughout, unless otherwise stated.

1D ME simulations tend to overestimate the polarizability. To keep the polarizability at a reasonable level, the
ionization potential had to be chosen slightly higher than usual values of complex materials (for instance
benzene: $\Ip=0.35$\,at.u.). The softening parameter used for the SAE system is $\aen=1.414$, and the parameters
to model highly polarizable atoms are $a_n=0.80$, $\aee=1.0$, and $Z=4$. The binding energy of the 4-electron
ground state is $\Enot{4} = 8.5$\,at.u. and the successive {\it single electron} ionization potentials are given
by 0.5, 1.07, 3.09 and 3.93\,at.u. The static polarizability is calculated by using the relation $\anot= 1/\sE
\int \Delta \rho(x) x \rmd x$, where $\Delta \rho$ is the change in electron density caused by the field $\sE$.
We find a polarizability of $\anot=31$\,\AA$^3$, which lies between the polarizability of transition metal atoms
and clusters, for example, Nb: $\anot=15$\,\AA$^3$, $\Ip=0.248$\,at.u.; C$_{60}$: $\anot=80$\,\AA$^3$,
$\Ip=0.279$\,at.u. Finally, the laser parameters are: center wavelength $\lambda = 1000$\,nm, peak intensity
$I = 2\ten{14}$\,W/cm$^2$, Gaussian envelope with FWHM width $\tau = 4 T_0$, and oscillation period
$T_0 = 3.33$\,fs.  The evolution of the wave function is calculated between $20$ optical cycles before and
$80$ optical cycles after the laser pulse maximum.

\begin{figure}[t]
\centerline{\includegraphics[width=0.8\columnwidth]{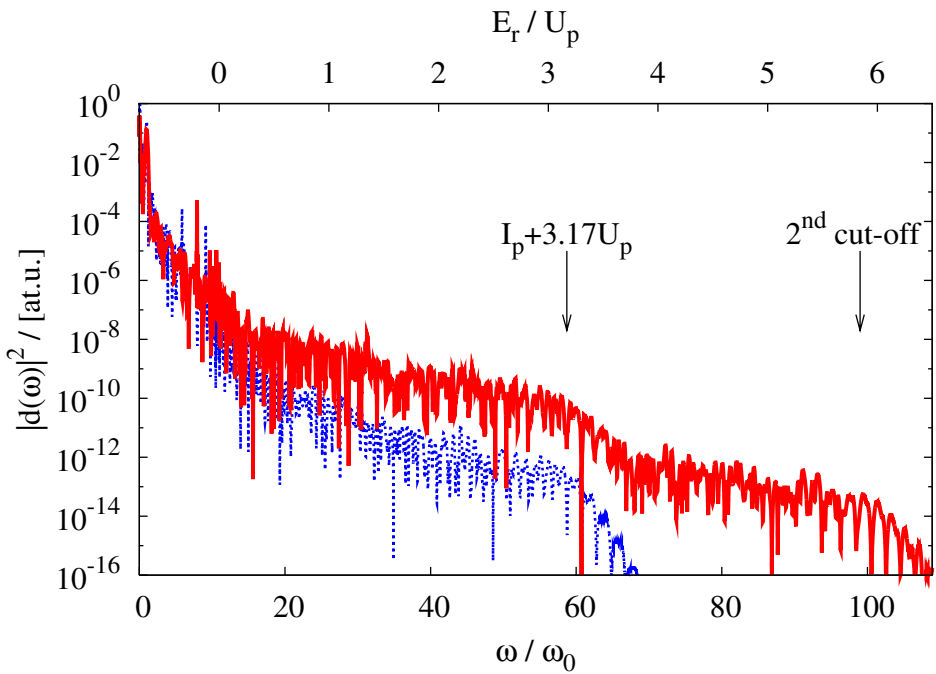}}
\caption{Spectra of the dipole moment squared, $\mid d(\omega) \mid^2$, of a highly polarizable ($\anot=31$\,\AA$^3$),
4-electron model-system, $\Ip=0.5$\,at.u., (thick full line), and corresponding SAE-calculation for the same $\Ip$
(thin dashed line). The lower x-axis gives the harmonic order, the upper x-axis gives the classical return energy,
$\Eret=(E-\Ip)/\Up$ with $E$ the harmonic photon energy. The standard cut off harmonic is at $(E-\Ip)/\Up=3.17$ and
is marked with an arrow. Laser parameters: $\lnt=1000$\,nm, peak intensity $I=2\ten{14}$\,W/cm$^2$, FWHM pulse
duration $\tau=4T_0$, optical period $T_0 = 3.33$\,fs, Gaussian envelope.
\label{fig:polhhg}}
\end{figure}
In figure~\ref{fig:polhhg} the harmonic spectrum is shown for a 4-electron system (full line) and a SAE system
(dashed line) with the same HOMO (highest occupied molecular orbital) ionization potential $\Ip = 0.5$\,at.u.
The harmonic spectrum is obtained as the modulus of the Fourier transform of the dipole expectation value,
$d(t)=\langle \Psi(t) | \sum_{i=1}^4 x_i | \Psi(t) \rangle$. $d(t)$ was sampled 256 times per optical cycle.
Note, however, that the time step in-between these sample points was self-adaptive. The cut off energy
$E= \Ip + 3.17 \Up$ is in agreement with the standard SAE cut off law \cite{corkum93,krause92}. Here
$\Up=(\sE_0/2 \onot)^2=0.68$\,at.u. is the ponderomotive energy, $\sE_0$ is the laser peak field strength, and
$\onot$ denotes the laser circular center frequency. The ME spectrum reveals in addition to the regular, first
cut off a second one. Note, that in the SAE system, ionization is saturated before the peak of the laser pulse,
which is not the case for the multi-electron case, where the saturation intensity is increased due to the
molecule's polarizability \cite{bhardwaj03,kitzler04}. Thus the early saturation of the SAE system reduces the
probability of electron trajectories that return with high energy and therewith results in a low high harmonic
yield of the plateau.

\begin{figure}[t]
\centerline{\includegraphics[width=\columnwidth]{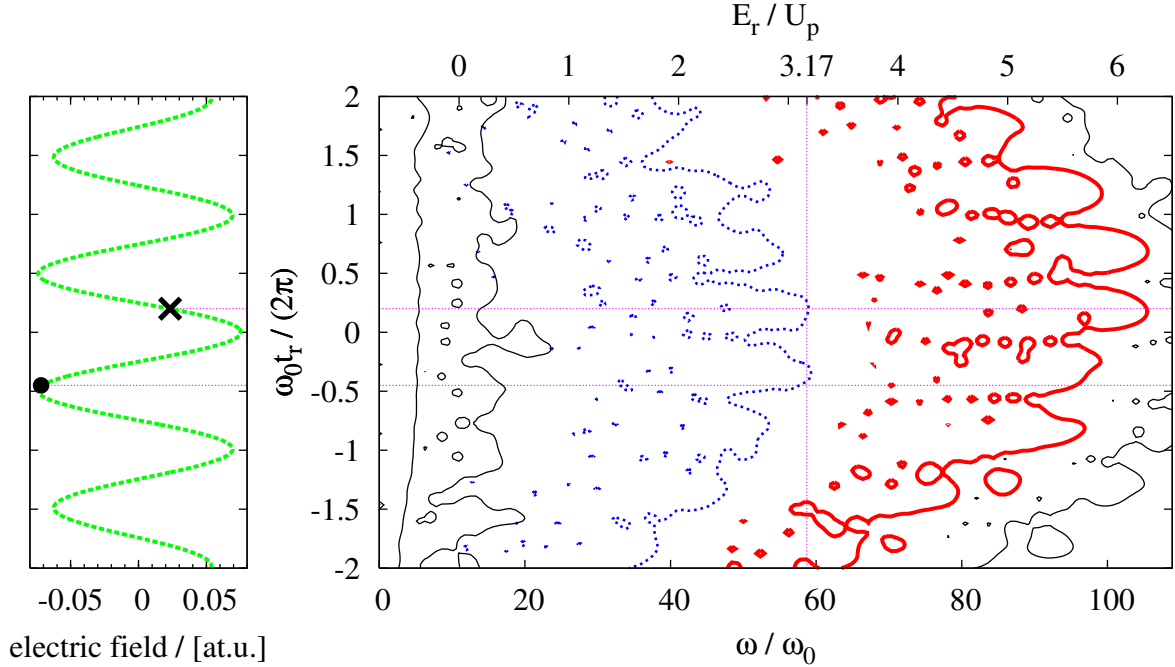}}
\caption{Contour plot of a time-frequency analysis of the 4-electron spectrum in Figure.~\ref{fig:polhhg}
(right panel). The window-function is a Gaussian pulse with 0.2 optical cycles FWHM duration. Contours differ
by a factor of $10^3$, decreasing from left to right. The return time $\tret$ is plotted versus the harmonic
frequency normalized to the laser frequency. Here the thick dashed line corresponds to the standard cut off,
while the thick full line represents the second cut off. The left panel shows the corresponding laser electric
field. The time of birth and the time of return for a classical electron acquiring the maximum kinetic
energy during its excursion in the laser field are marked by a dot and a cross, respectively.
\label{fig:polhhg-ttww}}
\end{figure}
To identify the origin of the second plateau we have performed a time-frequency analysis of the ME spectrum,
depicted in figure~\ref{fig:polhhg-ttww}. The dipole moment is truncated by the window function
$\sw (t;\Tw)=1/(\pi\Tw )^{1/4} \exp\LB-(t-t_r)^2/ (2\Tw^2)\RB$ with $\Tw= 0.2T_0$ and then Fourier transformed, i.e.
$\tilde{d}(\omega,\tret;\Tw)= \mathscr{F}\{\sw (\tret;\Tw)d(\tret)\}$. The harmonics corresponding to the first
and second plateau are depicted by the thick, dashed, and the thick, full lines, respectively. The time-
frequency analysis is a way to connect the quantum mechanical result with the classical three-step model
\cite{corkum93} model of HHG. It cuts small chunks in time out of the wavefunction and determines their energy at
the time of return to the parent ion. For a SAE harmonic spectrum, the resulting graph of harmonic energy versus
return time is very close to the result obtained by the classical three-step analysis. This correspondence allows
an interpretation of the the time-frequency plot and of HHG in terms of classical trajectories.

The thick dashed line in figure~\ref{fig:polhhg-ttww} denotes the regular, first cut off. A comparison to the
SAE three-step model allows us to determine the importance of ME effects in HHG. The electron return phase of
the cut off trajectory creating the highest harmonic in figure~\ref{fig:polhhg-ttww} is around 60\dg after the pulse
maximum. This is shifted with respect to the three-step model cut off trajectory that is born at 163\dg before
and returns at 80\dg after the pulse maximum \cite{corkum93} (see left panel in figure~\ref{fig:polhhg-ttww}).
The difference in the return times arises from a many-body effect. In ME systems the laser field induces a
polarization of the bound electrons that exerts a repelling force on the ionizing electron. This additional
potential decreases rapidly with the distance from the parent system and hence, affects the electron trajectories
only in the vicinity of the nucleus. Therefore, it mostly shifts the birth and return time of the electron
trajectories and only weakly affects the highest achievable cut off energy.

The thick full line in figure~\ref{fig:polhhg-ttww} corresponds to the second cut off. Surprisingly, the first
and second plateau show similar patterns. The maximum energy in each half cycle occurs at the same return phase for
both plateaus. This strongly indicates that the harmonics in both plateaus are generated by the \emph{same} electron
trajectories, starting from different initial states. The strong laser field brings the medium in a coherent
superposition of ground and excited states. HHG from excited states can take place as long as the phase of ground
and excited states are coherently locked.

In ME systems there exist two types of excitation, collective excitations and individual particle excitations.
Single electron excitations can be excluded for the following reasons: (i) The SAE calculation in
figure~\ref{fig:polhhg} does not show a second plateau. (ii) The energy difference between the first and second cut
off is larger than the HOMO ionization potential. (iii) The absence of doubly ionized states excludes HHG from a
deeper bound electron. (iv) While HHG from a coherent superposition of the ground state and an excited single-electron
state does indeed produce two plateaus, it does not result in an over all increase of the standard cut off law
\cite{watson96}, because the HOMO electron is born with the same energy after ionization regardless of its initial
state.

\begin{figure}[t]
\centerline{\includegraphics[width=\columnwidth]{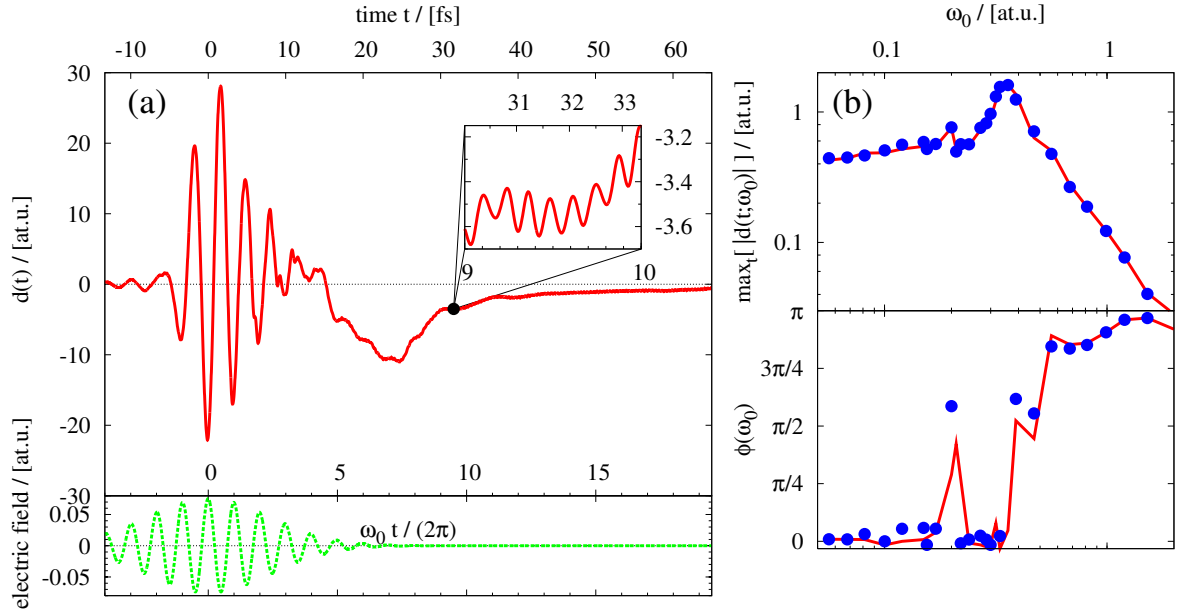}}
\caption{(a) Time dependence of the applied laser electric field (lower panel) and the resulting dipole moment,
$d(t)$ (upper panel), of the highly polarizable molecule in figure \ref{fig:polhhg}. In the inset we show a
magnified part of the dipole moment to illustrated the remaining excitations after the end of the laser pulse.
(b) Maximum excursion of the center of gravity of the electron density (upper panel) and the phase shift, $\phi$,
between the dipole signal and the laser electric field (lower panel) as a function of the applied laser frequency.
The full line is obtained for $m=8$ while the dots correspond to $m=12$ calculations. The closeness of agreement
for both sets of calculations confirms that the electronic structure of the model molecule is well approximated
and essentially converged. In (b) we have used a continues-wave laser which was linearly switched on over 4
optical cycles reaching a maximum intensity of $2\ten{13}$\,W/cm$^2$. The pulse was propagated over 13 optical
cycles.
\label{fig:polhhg-tt}}
\end{figure}

The excitation mechanism responsible for the second plateau is revealed in figure~\ref{fig:polhhg-tt}. In
sub-figure~\ref{fig:polhhg-tt}(a) the dipole moment, $d(t)$, is plotted as a function of time. We find that
the dipole moment exhibits oscillations even after the laser pulse, proving a laser induced excitation of
the system. The close-up in the inset of figure~\ref{fig:polhhg-tt}(a) shows that the excitation oscillates
at a frequency $\omega_p = 0.35$\,at.u. In figure~\ref{fig:polhhg-tt}(b) the excitation spectrum of the
system is determined by probing the response of the system to a plane wave laser signal as function of the
laser center frequency $\omega_0$. The maximum excursion of the center of gravity of the electron density
and the phase shift $\phi$ between dipole signal, $d$, and laser electric field are plotted. At resonance,
light absorption is maximum, and the center of charge motion of the electron is approximately $90$\dg out of
phase with the laser field.

The resonance in figure~\ref{fig:polhhg-tt}(b) is a collective plasmon resonance.
In 3D plasmas and clusters the collective excitation of the bound electrons is referred to as a plasma-wave
and as a plasmon, respectively. We define here the term plasmon as the corresponding collective motion of the
bound electrons of our 1D model system. The collective frequency depends on the system geometry, explaining the
difference between plasma and plasmon frequency. As our model system is neither the 1D limiting case of a bulk nor
of a sphere, the usual 3D plasma/plasmon frequency is not applicable. The 1D plasmon frequency is determined by
the 1D geometry of our model system. As an analytical expression is currently not known, we use the above
determined numerical value.

Single electron excitations cannot explain the resonance in figure~\ref{fig:polhhg-tt}. First because they have a
narrow linewidth. The broad width ($\approx 0.1$\,at.u.) of the resonance is a strong indicator of a collective
process. Second, the oscillation shown in figure~\ref{fig:polhhg-tt}(a) decays. The decay is also a typical
signature of a plasmon, as due to microscopic collisions energy is transferred from the collective electron motion
into thermal, single electron motion. In contrast to that, the life time of a single-electron excitation is
infinite. We have tested that the decay is not an artifact of the MCTDHF formalism. Increasing the number of
determinants does not change the time-dependence of the dipole signal significantly. Also, this decay does
not come from ionization and a decrease of the norm at the absorbing boundaries. We find that the ionization yield
is virtually constant (increasing by $0.003$ during the last $40$ optical cycles of the simulation time), while
the amplitude of the plasmon oscillation is reduced by almost a factor of two.

The match between the plasmon frequency of $\omega_p=0.35$\,at.u. and the frequency of the dipole oscillation
in figure~\ref{fig:polhhg-tt}(a) proves that the plasmon excitation is responsible for the second cut off
observed in HHG. Moreover, further analysis shows that multi-plasmon states are responsible for the second
plateau in the high harmonic signal (figure~\ref{fig:polhhg}). The oscillation in figure~\ref{fig:polhhg-tt}(a)
has non-sinusoidal components. A Fourier transform of $d(t)$ shows frequency components at multiples of the
plasmon frequency, $k \op$, $k=1,2,3,...$. Hence, the non-sinusoidal behavior arises from the interference
between these multi-plasmon states, quivering at multiples of the plasma frequency. Consequently, the width
of the second harmonic plateau is determined by the highest order of the excited multi-plasmon state, which
is $k=4$ in figure~\ref{fig:polhhg}. The multi-plasmon excitation comes from the nonlinear (anharmonic)
part of the binding potential. Whereas, close to the center the potential has a quadratic (harmonic oscillator)
space dependence supporting a single plasmon, the far-range Coulomb part of the potential adds nonlinear terms
responsible for the creation of multiple harmonics of the plasmon oscillation.

In contrast to single electron systems, in ME systems the collective excitation energy adds to the harmonic
cut off. The reason is that the collective energy stays in the remaining bound electrons and does not get
lost while the valence electron makes its excursion into the continuum. To elucidate this point we define
the ground state energies of the neutral and singly ionized system, $E_0^{(4)}$ and $E_0^{(3)}$, respectively.
The energies of the according plasmon states are given by $E_0^{(4)}+\op$ and $E_0^{(3)}+\op$. Here, we
neglect the difference in the plasmon frequencies between the neutral and the singly ionized state, since
the difference is of the order of the difference between two adjacent harmonics. As a result, in both cases
the HOMO potential is given by $\Ip= E_0^{(4)} - E_0^{(3)}$. Although for our ME system the HOMO ionization
potential for the plasmon state is slightly smaller, this is a reasonable approximation. In particular, since
the difference further decreases for an increasing number of electrons and will eventually disappear in real
ME system which usually have considerably more than four electrons.

\begin{figure}[t]
\centerline{\includegraphics[width=0.8\columnwidth]{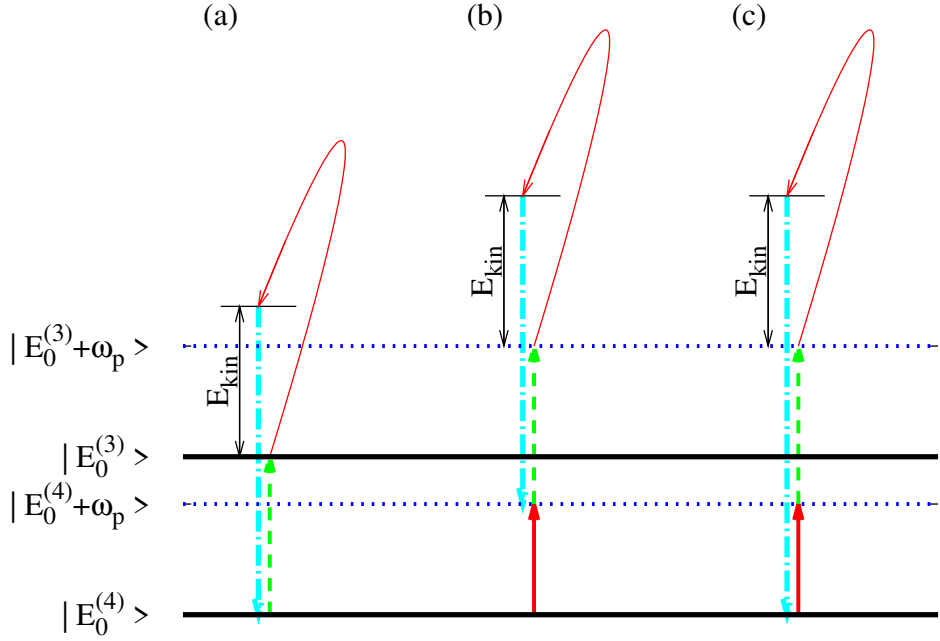}}
\caption{Energy level diagram for different pathways for HHG in
  polarizable molecules. (a) standard HHG: an electron is set free by tunnel
  ionization (broken arrow), quivers in the electric laser field (thin full
  arrow) gaining kinetic energy, $E_{\mbox{\scriptsize kin}}$, and recombines
  to the 4-electron ground state, $\Enot{4}$ (dashed dotted arrow). (b) as (a)
  but starting from and returning to the plasmon state $E_0^{(4)}+\op$. (c) as
  (b) but returning to the ground state, $\Enot{4}$.\label{fig:hhgpath} Here,
  $\Enot{3}$ and $E_0^{(3)}+\op$ denote the singly ionized ground state and
  singly ionized plasmon state, respectively.}
\end{figure}
There are different pathways by which HHG can take place, which are 
illustrated in figure~\ref{fig:hhgpath}. Before ionization the system is in its 
ground state, $\Enot{4}$, remains in the ground state, $\Enot{3}$, after 
ionization of the valence electron and returns upon recombination to its initial 
state [figure~\ref{fig:hhgpath}(a)]. This is the standard HHG situation. 
The system may also start out in a plasmon state, $\Enot{4}+\op$, remains in the 
plasmon state, $\Enot{3}+\op$, after ionization, and returns to its four 
electron plasmon state upon recombination [figure~\ref{fig:hhgpath}(b)]. 
For both cases the cut off law is $3.17\Up+[\Enot{4} -\Enot{3}] = 3.17\Up+\Ip$ 
since for the latterthe plasmon frequency chancels out. However, if ionization 
starts from the plasmon state, $E_0^{(4)}+\op$, but the electron returns to the 
ground sate, $\Enot{4}$, upon recombination, the plasmon energy is converted 
into harmonic photon energy, extending the cut off, i.e. $3.17\Up+\Ip+\op$ 
[figure~\ref{fig:hhgpath}(a)].

So far we have discussed the single system response. At the moment, the observation of HHG from excited states is experimentally untested. A significant macroscopic signal will only be created, when the individual systems emit harmonic radiation in phase. Hence, the question has to be answered whether the second plateau can be detected in a macroscopic medium consisting of many individual systems. We believe that this is the case for the following reason. The phase of the harmonic signal emitted by a single system is determined by
the phase difference between ground state and excited (plasmon) state. In order for coherent emission to occur, the phase difference between ground and excited state has to be the same for all emitters. As the plasmon excitation is driven by the laser field, the phase difference between ground and (plasmon) excited state is exclusively determined by the laser field, and therefore is the same for each system. As a result, the multi-plasmon states are (laser) phase locked to the ground state. The contributions from individual atoms add up coherently and HHG can take place from the ground as well as from excited states.

Finally, with respect to experimental observation, we believe that metal clusters with their simple geometry and with plasmon energies of a couple of eV are good candidates. Although atoms also support collective oscillations, which are known as giant (shape) resonances \cite{amusia00}, their life time is usually very short, 3-4 times the plasmon oscillation period. As a result the plasmon will likely decay before the active electron can return and create harmonic radiation.

\section{Conclusion}

High harmonic generation (HHG) in complex multi-electron (ME) systems was investigated within the framework of
the multi-configuration time-dependent Hartree-Fock (MCTDHF) method. Our analysis of HHG spectra in complex
multi-electron systems revealed two plateaus. The first cut off agrees with the cut off law of noble gases. The
second plateau is generated due to non-linear excitation of collective
plasmon oscillations. It arises from electrons that are ionized from an excited plasmon state and recombine to
the ground state. The ratio of first to second plateau determines the population of the plasmon-states. From
the extension of the second plateau the order of the highest excited plasmon state can be inferred. Thus, the
here identified plasmon signature presents a novel tool for the investigation of the non-perturbative
multi-electron dynamics in complex materials, a regime that is experimentally very difficult to access otherwise.

\section*{References}
\bibliography{zang1208}

\end{document}